\documentclass[prd,aps,twocolumn,nofootinbib,preprintnumbers,superscriptaddress,preprintnumbers,balancelastpage,longbibliography]{revtex4-2}
\usepackage[english]{babel}

\usepackage{amsmath,mathtools,physics,xfrac}
\usepackage{graphicx}
\usepackage{afterpage}
\usepackage{float}
\usepackage{subfigure}
\usepackage{rotating}
\usepackage{multirow}
\usepackage{tabularx}
\usepackage{booktabs}
\usepackage{fancyhdr}
\usepackage[utf8]{inputenc}
\usepackage{theorem}
\usepackage{moreverb}
\usepackage{euscript}
\usepackage{psfrag}
\usepackage{slashed}
\usepackage{mathtools}
\usepackage{makecell}
\usepackage{adjustbox}
\usepackage{dcolumn}
\usepackage{bm}
\usepackage[dvipsnames]{xcolor}
\usepackage{graphics}
\usepackage{hyperref}
\usepackage{chngcntr}
\usepackage{aas_macros}
\usepackage{tabularx}
\usepackage{color,xcolor}
\usepackage{gensymb}
\usepackage[normalem]{ulem}
\hypersetup{
     pdftitle={Better Call LHAASO},
     colorlinks   = true,
     citecolor    = Green,
     urlcolor     = Green,
     linkcolor    = Green
}

\definecolor{darkblue}{rgb}{0.0,0.0,0.75}
\definecolor{darkred}{rgb}{0.6,0.0,0}
\definecolor{darkgreen}{rgb}{0.0,0.6,0.}

\counterwithout{equation}{section}

\usepackage{tikz,xcolor,hyperref}

\definecolor{lime}{HTML}{A6CE39}
\DeclareRobustCommand{\orcidicon}{\hspace{-1mm}
	\begin{tikzpicture}
		\draw[lime, fill=lime] (0,0) 
		circle [radius=0.16] 
		node[white] {{\fontfamily{qag}\selectfont \tiny \,ID}};
		\draw[white, fill=white] (-0.0525,0.095) 
		circle [radius=0.007];
	\end{tikzpicture}
	\hspace{-3mm}
}

\foreach \x in {A, ..., Z}{\expandafter\xdef\csname orcid\x\endcsname{\noexpand\href{https://orcid.org/\csname orcidauthor\x\endcsname}
		{\noexpand\orcidicon}}
}


\begin{document}



\title{\boldmath Too Heavy to Hide: Gamma-Ray Constraints on Annihilating Dark Matter beyond Unitarity}

\author{Abhishek Dubey\orcidA{}}
\email{abhishekd1@iisc.ac.in}
\affiliation{Centre for High Energy Physics, Indian Institute of Science, C.\,V.\,Raman Avenue, Bengaluru 560012, India}

\author{Deep Jyoti Das\orcidB{}}
\email{deepjyoti@iisc.ac.in}
\affiliation{Centre for High Energy Physics, Indian Institute of Science, C.\,V.\,Raman Avenue, Bengaluru 560012, India}

\author{Akash Kumar Saha\orcidC{}}
\email{akashks@iisc.ac.in}

\affiliation{Centre for High Energy Physics, Indian Institute of Science, C.\,V.\,Raman Avenue, Bengaluru 560012, India}

	
	
	\begin{abstract}

    The measurement of high energy diffuse gamma rays by various ground-based air shower detectors have opened a new chapter for high energy particle physics and astrophysics. The broad range of viable dark matter candidates motivates extending indirect searches to heavier dark matter masses, opening new opportunities to uncover the nature of dark matter. If dark matter is composite rather than point-like, then the thermal unitarity bound can be relaxed, opening up the possibility of dark matter masses far beyond the electroweak scale. We perform a model agnostic search for heavy annihilating dark matter using the gamma-ray measurements and upper limits from Tibet AS$_\gamma$, LHAASO, KASCADE-Grande, Pierre Auger Observatory, and Telescope Array. These highest energy datasets enable us to probe new regions of parameter space and set world-leading limits on the annihilation cross sections for dark matter masses $10^5$--$10^{12}$ GeV. Our work highlights the power of high energy gamma-ray datasets in discovering heavy dark matter signatures in the near future.

	\end{abstract}
	
	\maketitle
	
\section{\label{sec:level1}Introduction}
The non-gravitational nature of dark matter (DM) remains one of the most important mysteries of modern physics\,\cite{Bertone:2016nfn,Cirelli:2024ssz, Strigari:2012acq, Lisanti:2016jxe, Lin:2019uvt}. All the existing evidences including galaxy rotation curves\,\cite{Rubin:1970zza}, gravitational lensing\,\cite{Clowe:2006eq}, cosmic microwave background anisotropies\,\cite{Planck:2018vyg} reveal the gravitational nature of DM.  The possible mass range of interest for DM spans almost 90 orders of magnitude. This warrants a model agnostic and careful search for DM with all sorts of existing datasets.

In the standard freeze-out paradigm, DM particles were in thermal equilibrium with Standard Model (SM) particles in the early universe. As the Universe expands and cools down, DM particles get decoupled from the SM bath and make up the present DM density. In the present universe within DM dense regions, DM particles can annihilate to SM final particles which can be detected with existing telescopes. This search technique is known as the indirect detection of DM. Traditional Weakly Interacting Massive Particle (WIMP) searches have focused on GeV-TeV DM masses. This is partly motivated by the fact that in these masses, WIMPs can satisfy the correct relic abundance without additional assumptions. Lately, partly due to the null results from various WIMP searches from both ground and space-based detectors, alternate DM models spanning both lighter and heavier masses are gaining attention. In particular, various Beyond the Standard Model (BSM) models predict heavy DM masses ranging from PeV masses to Planck scale masses that can make up the total DM abundance in the present universe\,\cite{Carney:2022gse}. 

\begin{figure}[t]
\centering
\includegraphics[width=\columnwidth]{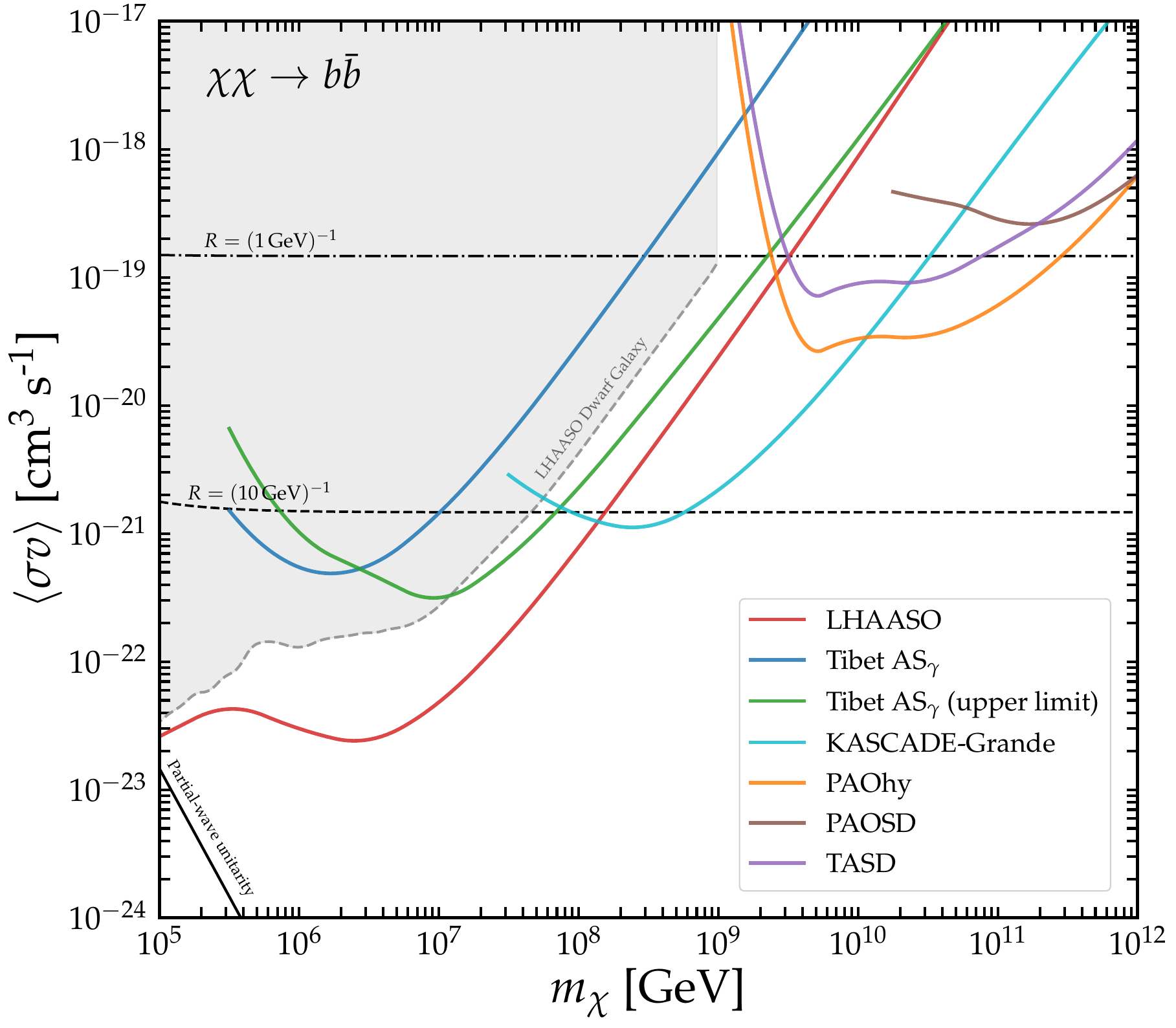}
\caption{Constraints on DM annihilation cross section, $\langle\sigma v\rangle$ as a function of DM mass $m_\chi$ for the annihilation channel $\chi \chi\rightarrow b \bar{b}$ (100\% branching ratio) using different gamma-ray datasets. The previous limits in the parameter space is taken from Ref.\,\cite{LHAASO:2024upb} (grey shaded region). The black solid, dashed, and dot-dashed curves show the partial-wave unitarity and composite unitarity for DM radius of (10 GeV)$^{-1}$ and (1 GeV)$^{-1}$, respectively (see Sec.\,\ref{unitarity} for details).  }
\label{fig:bbbar_constraints}
\end{figure}

High energy gamma rays offer a unique window to explore the Universe at the highest energies\,\cite{Moskalenko:2004vh, Kelner:2006tc,Kappes:2006fg,Abdo:2007ad,2009herb.book.....D,2012APh....35..503G,Lipari:2018gzn,Moskalenko:2005ng,1989ApJ...342..379W, Capanema:2020oet, 2010ApJ...710.1530V,Vernetto:2016alq,DeAngelis:2013jna,Ruffini:2015oha,Sudoh:2022sdk,1971NASSP.249.....S, Bose:2022ghr, Cardillo:2023hbb,Lipari:2024pzo,Ehlert:2023btz}. Cosmic-ray particles can interact with the interstellar medium gas and produce high-energy diffuse gamma rays. A wide range of ground- and space-based instruments have observed the high-energy gamma-ray sky at different energies. At the PeV-frontier, Tibet AS$_\gamma$\,\cite{TibetASgamma:2021tpz} and Large High Altitude Air Shower Observatory (LHAASO) have observed diffuse gamma-ray from the Galactic plane region\,\cite{LHAASO:2023gne}. In addition, various ground-based air-shower detectors like CASA-MIA\,\cite{CASA-MIA:1997tns}, KASCADE-Grande\,\cite{2017ApJ...848....1A}, Telescope Array (TA)\,\cite{TelescopeArray:2018rbt}, Pierre Auger Observatory (PAO)\,\cite{PierreAuger:2015fol, PierreAuger:2016kuz} have provided upper limits on ultra high energy (UHE) gamma rays ($\rm E_{\gamma}\gtrsim10^{15}\,eV$). These results provide detailed insights into the origin of UHE cosmic rays and their interactions.

In this work, we use the latest UHE gamma-ray measurements from air-shower detectors to set stringent constraints on heavy annihilating DM within the mass range $10^5 - 10^{12}$ GeV. In Fig.\,\ref{fig:bbbar_constraints},
we show our results for $\chi\chi\rightarrow b \bar{b}$ channel (100\% branching fraction). With the latest datasets and upper limits from experiments like Tibet AS$_\gamma$, LHAASO, TA, KASCADE-Grande, and PAO, we are able to probe new regions of parameter space for heavy annihilating DM. This makes upcoming observations by air-shower experiments an exciting avenue for discovering heavy annihilating DM. Previously, high energy gamma-ray, neutrino, and cosmic ray observations have provided stringent constraints on heavy decaying DM\,\cite{Ishiwata:2008cu, Murase:2012xs, Murase:2015gea, Esmaili:2015xpa, Esmaili:2014rma,Murase:2015gea, Kalashev:2016cre,Kalashev:2017ijd, Kachelriess:2018rty, Blanco:2018esa, Ishiwata:2019aet, Alcantara:2019sco, Cohen:2016uyg, Bhattacharya:2019ucd,Kalashev:2019xkw,  Kuznetsov:2016fjt,Sui:2018bbh,Chianese:2019kyl,Maity:2021umk,Anchordoqui:2021crl, Esmaili:2021yaw,Chianese:2021jke,IceCube:2022clp,LHAASO:2022yxw,Arguelles:2022nbl,Aramaki:2022zpw,Skrzypek:2022hpy,Hambye:2021moy,Allahverdi:2023nov,IceCube:2023gku,PierreAuger:2023vql,IceCube:2023ies, PierreAuger:2023vql, LHAASO:2024upb,Sarmah:2024ffy,Das:2024bed,Dekker:2019gpe,Ng:2020ghe,Deliyergiyev:2022bvp,Leung:2023gwp,Das:2023wtk,Munbodh:2024ast,Song:2024vdc,Liu:2025vpz,Murase:2025uwv,Berghaus:2025jwb,LAT:2025wdk,Feldstein:2013kka,Murase:2015gea,Kachelriess:2018rty, Dekker:2019gpe,Arguelles:2019ouk,Chianese:2021htv,Arguelles:2022nbl, Kohri:2025bsn, Aloisio:2025nts, Mukherjee:2025btq, Barman:2025gjr,Berghaus:2025jwb,Chianese:2026cfz,Rocamora:2025ddt}. 

This paper is structured as follows. In section \ref{unitarity}, we discuss heavy annihilating DM beyond the vanilla unitarity considerations. Section \ref{DM flux} gives an outline of the gamma-ray flux produced from heavy annihilating DM. Next, we discuss the observations used in this work in section \ref{dataset}. In section \ref{results}, we discuss our results, future prospects, and conclude.

\section{Beyond the thermal unitarity bound}
\label{unitarity}

Unitarity is an important consideration for determining the maximum mass of DM particles in the freeze-out production scenario. For DM masses above 194 TeV, thermal freezeout cannot generate the correct relic abundance. This comes from the fact that for s-wave annihilation, the total cross section $\langle\sigma v\rangle \propto 1/{m_\chi^2}$, where $m_\chi$ is DM mass and during freeze-out, the DM abundance $\Omega_\chi \propto 1/\langle\sigma v\rangle$. Therefore, as we increase the DM mass above a particular threshold, the annihilation rate drops below what is required to satisfy the correct relic abundance of DM. It was first pointed out in Ref.\,\cite{Griest:1989wd} which was later refined in Ref.\,\cite{Smirnov:2019ngs}. We also note that, in addition to the traditional thermal WIMP paradigm, DM searches have resulted in stringent bounds for a wide range of DM candidates, extending from fuzzy dark matter to primordial black hole dark matter\,\cite{Dalal:2022rmp, Saha:2025any,Boyarsky:2018tvu, Saha:2025wgg, Saha:2021pqf, Saha:2024ies,Murgia:2019duy}.

 Given the null results from collider experiments as well as from indirect searches by telescopes like Fermi-LAT, one might ask how strict is the thermal unitarity limit and how one can move past the conventional WIMP paradigm. In the following, we briefly mention the different ways in which the unitarity limit have been relaxed: the unitarity mass upper limit assumes that the DM particles were in thermal equilibrium with the surrounding SM bath and then achieved freeze-out at some point. This picture will be modified in the case where DM is part of a dark sector that was not in thermal contact with the SM bath in the early universe. These stable DM particles can then annihilate to lighter dark states which later can decay to SM particles. This scenario can circumvent the usual unitarity limit and can allow ultra heavy DM as a viable DM candidate\,\cite{Berlin:2016gtr, Berlin:2016vnh, Heurtier:2019eou, Davoudiasl:2019xeb}. If DM is long-lived and can scatter against another dark species particles which were in equilibrium with the SM bath, then again much heavier DM particles become possible\,\cite{Kramer:2020sbb}. In case of an early matter-domination after freeze-out, DM particles can annihilate with a smaller annihilation cross-section to avoid dilution and satisfy the relic. This can also relax the unitarity limit\,\cite{Bramante:2017obj,Cirelli:2018iax,Bhatia:2020itt}. For DM models where the DM mass is much heavier than the mediator, unitarity bound can be altered due to Sommerfeld enhancement of cross section and formation of bound state\,\cite{Hisano:2006nn,vonHarling:2014kha,Baldes:2017gzw,Smirnov:2019ngs,Bottaro:2021snn}. Ref.\,\cite{Kim:2019udq} showed that if DM consists of nearly degenerate particles that can scatter with the SM bath then the DM mass can even be pushed upto $10^{14}$ GeV. 

 If instead of being a point particle, DM is composite in nature, then DM particles can be ultra heavy and still obey the unitarity condition. In that case, each of the partial waves satisfies the unitarity limit and all of the partial waves contribute to the full annihilation cross-section. For $J^{\rm th}$ angular momentum, the annihilation cross-section must satisfy
\begin{eqnarray}
    \sigma^J \leq \frac{4\pi (2J+1)}{m_\chi^2v_{\rm}^2}\,\,,
\end{eqnarray}
where $v$ is the relative velocity between DM particles. The total annihilation cross section is then bounded by\,\cite{Griest:1989wd, Tak:2022vkb}
\begin{eqnarray}
    \sigma v &\leq&\,\frac{4\pi}{m_\chi^2v}\sum_{J=0}^{J_{\rm max}}(2J+1)\,\\
    &\simeq& \,\frac{4\pi}{m_\chi^2v}(1+m_\chi v\,R)^2\,\,,
\end{eqnarray}

In the above relation we have substituted $J_{\rm max}=m_\chi v R$, where $R$ is the radius of the composite DM particle. We note that, for $m_\chi v R \gg 1$, the annihilation cross section upper limit becomes independent of the DM mass. In our work, we remain model agnostic and use the above relation as a benchmark for the maximum annihilation cross-section allowed for different values of $R$. For the relative velocity, we assume the local DM velocity dispersion $v \sim 10^{-3} c$, as the benchmark. We note that the above equation in the limit of $R\rightarrow0$ becomes the unitarity limit for point particle, $\langle\sigma v\rangle\leq 4\pi/(m_\chi^2 v)$. This is the maximum present day annihilation cross-section allowed today for point particle. This we refer to as `Partial-Wave Unitarity'.

\section{Calculating the gamma ray flux} 
\label{DM flux}
In this section, we outline the calculation of the total gamma-ray flux from DM annihilation. An important ingredient for this is the prompt gamma-ray spectra per DM annihilation, $\frac{dN}{dE_\gamma}\small(E_\gamma\small)$, that depends on the channel through which DM annihilates. We remain agnostic of the particle physics model and assume that DM annihilates to various two-body final states with 100\% branching ratio. The prompt 
 gamma-ray spectra produced from DM annihilation include Final State Radiation (FSR) and Electroweak corrections. For our work, we use the publicly available code, \texttt{HDMSpectra}\footnote{\href{}{\textcolor{purple}{https://github.com/nickrodd/HDMSpectra}}}\,\cite{Bauer:2020jay} to calculate the total gamma-ray spectra taking into account the primary and secondary contributions.

In addition to prompt photons, prompt electron- positrons produced in each annihilating channel can also give rise to gamma rays via inverse Compton scattering (ICS). This, however, is complicated by the fact that electrons lose energy through synchrotron radiation in the Galactic magnetic field.
The extragalactic gamma ray flux is calculated in a very similar fashion, except that one needs to take into account the redshifting of photons.  Absorption of high energy photons by background photons through pair-production and subsequent regeneration of upscattered photons by ICS gives rise to an electromagnetic cascade. For the Galactic and the Extragalactic DM annihilation, we include the Subhalo Boost Factor and the Extragalactic Clumping Factor, respectively. Below we outline the calculations for the various components of the total gamma-ray flux.

\subsection{Galactic prompt gamma-ray flux from DM annihilation}

In case of DM annihilation, the differential prompt gamma-ray flux is given by
\begin{equation}
\frac{d^2\Phi}{dE_\gamma \, d\Omega}(l,b)
=
\frac{\langle \sigma v \rangle}{8\pi \, m_{\chi}^2}
\int_0^\infty ds \,
\rho_\chi^2(s)\,
e^{-\tau(E_\gamma,s)}\frac{dN}{dE_\gamma}\,\,,
\label{eq: Annihilation flux}
\end{equation}
where $\langle \sigma\, v \rangle$ is the velocity averaged cross-section and 
$\frac{dN}{dE_\gamma}$ is the prompt gamma ray spectrum, $\tau(E_\gamma,s)$ is the optical depth due to the cosmic microwave background (CMB), starlight (SL), and infrared (IR) backgrounds\,\cite{Esmaili:2015xpa, Lipari:2018gzn,Galanti:2019rnl}, and $s$ is the line of sight (l.o.s.) distance. In the above equation, $\rm \rho_{\chi}$ is the DM density profile, which, as a benchmark, we take to be the Navarro-Frenk-White (NFW) profile\,\cite{Navarro:1996gj}
\begin{equation}
\rho_\chi(r)
=
\frac{\rho_s}
{\dfrac{r}{r_s} \left(1 + \dfrac{r}{r_s}\right)^2}\,\,,
\end{equation}
with $r_s=20\,\mathrm{kpc}$ and $\rho_s=0.318\,\mathrm{GeVcm^{-3}}$. The distance from the Galactic center, $r$, is related to $s$ by
\begin{equation}
r(s,b,l)
=
\sqrt{
R_\odot^2
- 2 s R_\odot \cos b \cos l
+ s^2
}\,\,,
\end{equation}
where $R_\odot = 8.3$ kpc is the distance of the Sun from the Galactic center, $b$ and $l$ are the Galactic latitude and longitude, respectively.

\subsection{Gamma ray flux from the ICS }

In addition to the prompt gamma rays, prompt electron-positrons can also be emitted for the channels under consideration.  These electrons and positrons can then upscatter the background low-energy photons to produce high-energy gamma rays. The differential flux of photons from the ICS of the prompt electron and positron is given by
\begin{equation}
\frac{d^2\Phi_{\mathrm{ICS}}}{dE_\gamma \, d\Omega}(l,b)
=
\frac{1}{4\pi\,E_\gamma}
\int_{\mathrm{los}} ds \,
j(E_\gamma,s) \, e^{-\tau(E_\gamma,s)}\,\,,
\end{equation}

where $j(E_\gamma,s)$ is the emissivity given by
\begin{equation}
j(E_\gamma,s)
=
2 \int_{m_e}^{m_{\rm \chi}}
dE_e \,
P_{\rm ICS}(E_\gamma,E_e,s)
\,
\frac{dn_{e^-}}{dE_e}(s,E_e) \, .
\end{equation}

In the above equation, $P_{\rm ICS}(E_\gamma, E_e,s)$ is the power emitted into photons of energy $E_\gamma$ by an electron with energy $E_e$, and $\frac{dn_{e^-}}{dE_e}(s, E_e)$ is the number density of electrons along the l.o.s. The factor of 2 in front accounts for ICS by both electrons and positrons. The number density of electron-positrons from DM annihilation is given by 

\begin{align}
\frac{dn_{e^-}}{dE_e}(s,E_e)
&=
\frac{\langle \sigma v \rangle \,\rho_{\rm\chi}^2(s)}
     {2\, b(E_e,s)\, m_{\rm \chi}^2}
\nonumber \\
&\quad \times
\int_{E_e}^{m_{\rm \chi}}
dE_s \,
\frac{dN_{e}}{dE_s}(E_s)
\, I_{\rm diff}(E_e,E_s,s) .
\end{align}
Here,  $\frac{dN_e}{dE_s}(E_s)$ is the spectrum of electrons produced from DM annihilation,
$b(E_e,s)$ is the energy loss rate, and $I_{\rm diff}(E_\gamma, E_e,s)$ is the diffuse halo function along l.o.s.  At higher energies, only ICS and synchrotron losses are important, thus the energy loss rate becomes
\begin{equation}
b(E_e,\vec{r})
=
- \frac{dE_e}{dt}
=
b_{\rm IC}(E_e,\vec{r})
+
b_{\rm syn}(E_e,\vec{r})\,,
\end{equation}
where the terms on the right-hand side are energy loss rates due to IC scattering and synchrotron, respectively. For discussions of $P_{\rm ICS}(E_\gamma,E_e,s,l,b)$ and $I_{\rm diff}(E_\gamma, E_e,s,l,b)$, we refer the readers to Ref.\,\cite{Dubey:2025ouh}, and the references therein.

\begin{figure*}
	\begin{center}
		\includegraphics[width=\columnwidth]{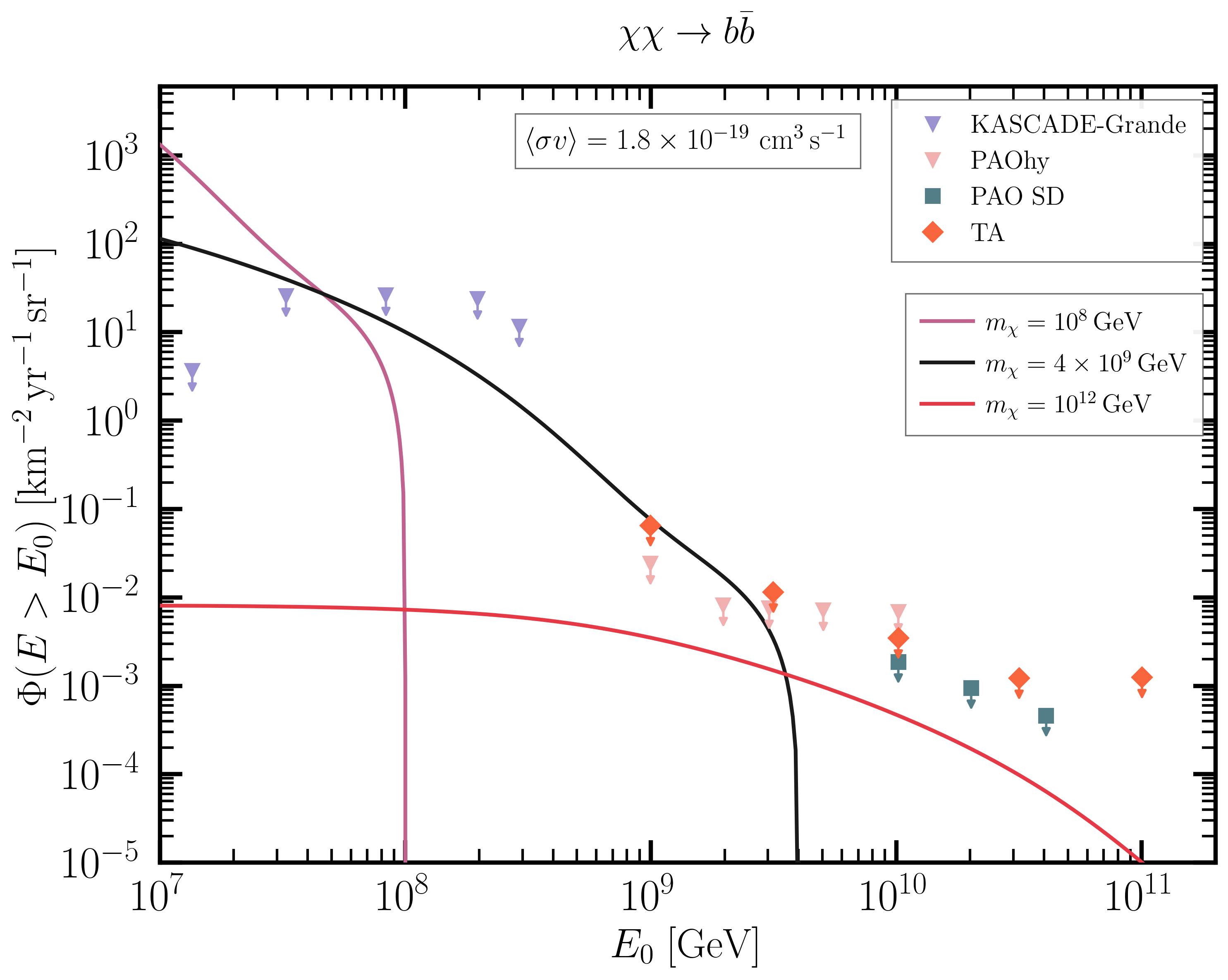}~~
		\includegraphics[width=\columnwidth]{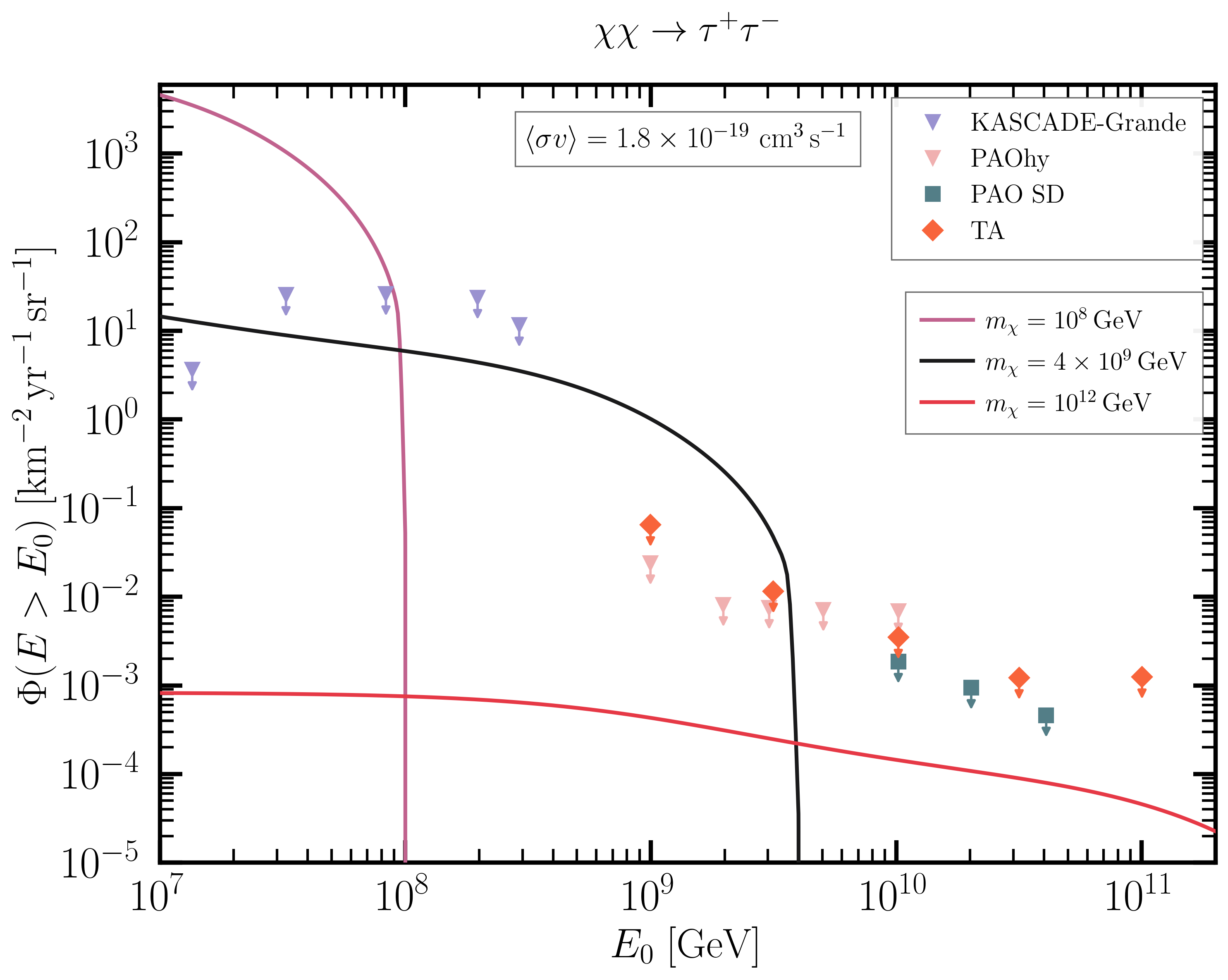}~~\\	
		\caption{Gamma-ray flux $\Phi$ integrated above energy $E_0$, denoted by $\Phi(E>E_0)$, as a function of $E_0$ for the DM annihilation channels $\chi \to b\bar{b}$ (\textbf{left panel})  and $\chi \to \tau^+\tau^-$ (\textbf{right panel}). In both the cases, we have used the benchmark values $\langle\sigma v\rangle = 1.8 \times 10^{-19}\,\rm cm^3 \rm s^{-1}$}
		\label{fig:flux_plots}
	\end{center}	
\end{figure*}

\subsection{Subhalo Boost Factor}
\label{subhaloboost}

DM annihilation rate scales as $\rho^2_{\chi}$, and in the CDM 
framework, the Milky Way halo contains a large population of subhalos spanning masses from $\sim 10^{-6}\,M_\odot$ (for neutralino-like 
candidates) up to a significant fraction of the host mass\,\cite{Bringmann:2009vf}. Since 
$\langle \rho^2_\chi \rangle \geq \langle \rho_\chi \rangle^2$, 
this clumpy substructure boosts the total annihilation luminosity above 
what the smooth NFW profile alone would give, which we denote by $L_{\text{host},0}$. We account for the enhanced luminosity through 
the subhalo boost factor, defined as the ratio of the total annihilation 
luminosity in subhalos to that of the smooth host halo~\cite{Ando:2019,Ng:2013xha}:
\begin{equation}
    B_\text{sh}(M, z) = \frac{1}{L_{\text{host},0}(M,z)} 
    \int dm\, \frac{dN_\text{sh}(m \mid M, z)}{dm}\, L_\text{sh}(m)
    \label{eq:bsh_def}
\end{equation}
where $m$ is the mass of the subhalo, $M$ is the mass of the Halo, hosting the subhalos and $z$ is the redshift under consideration. The presence of substructure modifies $\rho^2_\chi$ in Eq.~\eqref{eq: Annihilation flux} 
as
\begin{equation}
    \rho^2_\chi \;\rightarrow\; 
    \left(1 - f^2_\text{sh} + B_\text{sh}\right)\rho^2_\chi,
\end{equation}
where $f_\text{sh}$ is the fraction of the host mass contained in subhalos. 
For a Milky Way-mass halo at $z=0$, $f_\text{sh} \lesssim 0.1$--$0.2$ 
\cite{Ando:2019}, so the $f^2_\text{sh}$ correction is small and the flux 
in Eq.~\eqref{eq: Annihilation flux} is simply rescaled as
\begin{equation}
    \frac{d^2\Phi_\text{eff}}{dE_\gamma\,d\Omega} \;\approx\; 
    \left(1 + B_\text{sh}\right)
    \frac{d^2\Phi}{dE_\gamma\,d\Omega}.
    \label{eq:flux_boosted}
\end{equation}

We compute $B_\text{sh}$ using the semi-analytic model of 
Hiroshima, Ando \& Ishiyama (2018)~\cite{Hiroshima:2018} (H18), 
as reviewed and parametrized in Ando et al.\ (2019)~\cite{Ando:2019} 
(A19). The H18 model evolves the subhalo population from accretion 
to the present day using the extended Press-Schechter formalism 
combined with a tidal stripping prescription calibrated against 
$N$-body simulations, covering host masses $10^{-4}$\,--\,$10^{14}\,M_\odot$ 
and redshifts up to $z \sim 10$. Unlike earlier estimates that 
extrapolated power-law concentration-mass relations to unresolved 
scales yielding $B_\text{sh} \sim \mathcal{O}(100)$ for Milky 
Way-mass halos~\cite{Springel:2008,Gao:2012}, the H18 model 
accounts for the physical tidal evolution of subhalo concentrations, 
giving considerably more modest predictions of 
$B_\text{sh} \sim \mathcal{O}(1)$ for Milky-Way size hosts.

\begin{figure*}
	\begin{center}
		\includegraphics[width=\columnwidth]{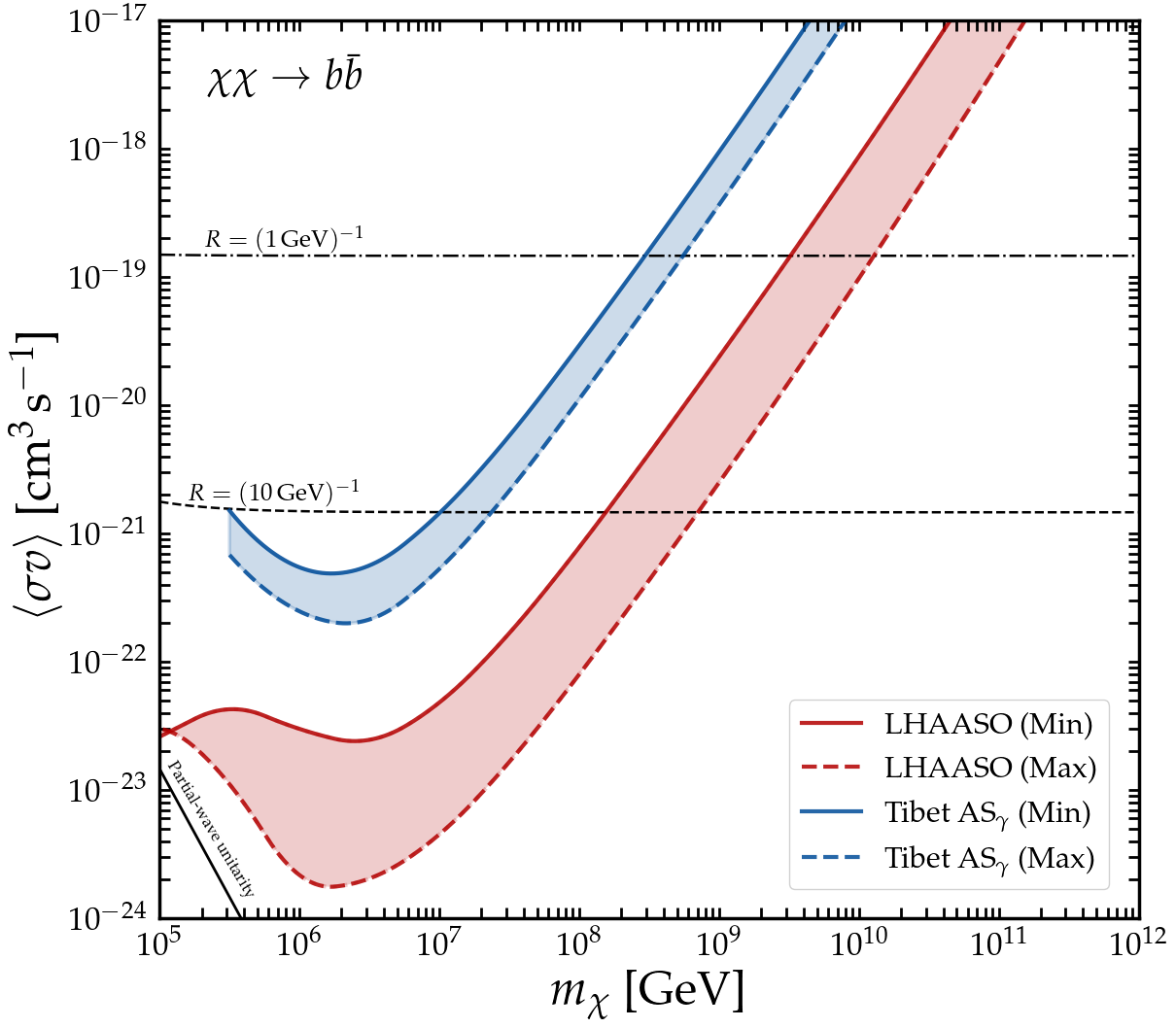}~~
		\includegraphics[width=\columnwidth]{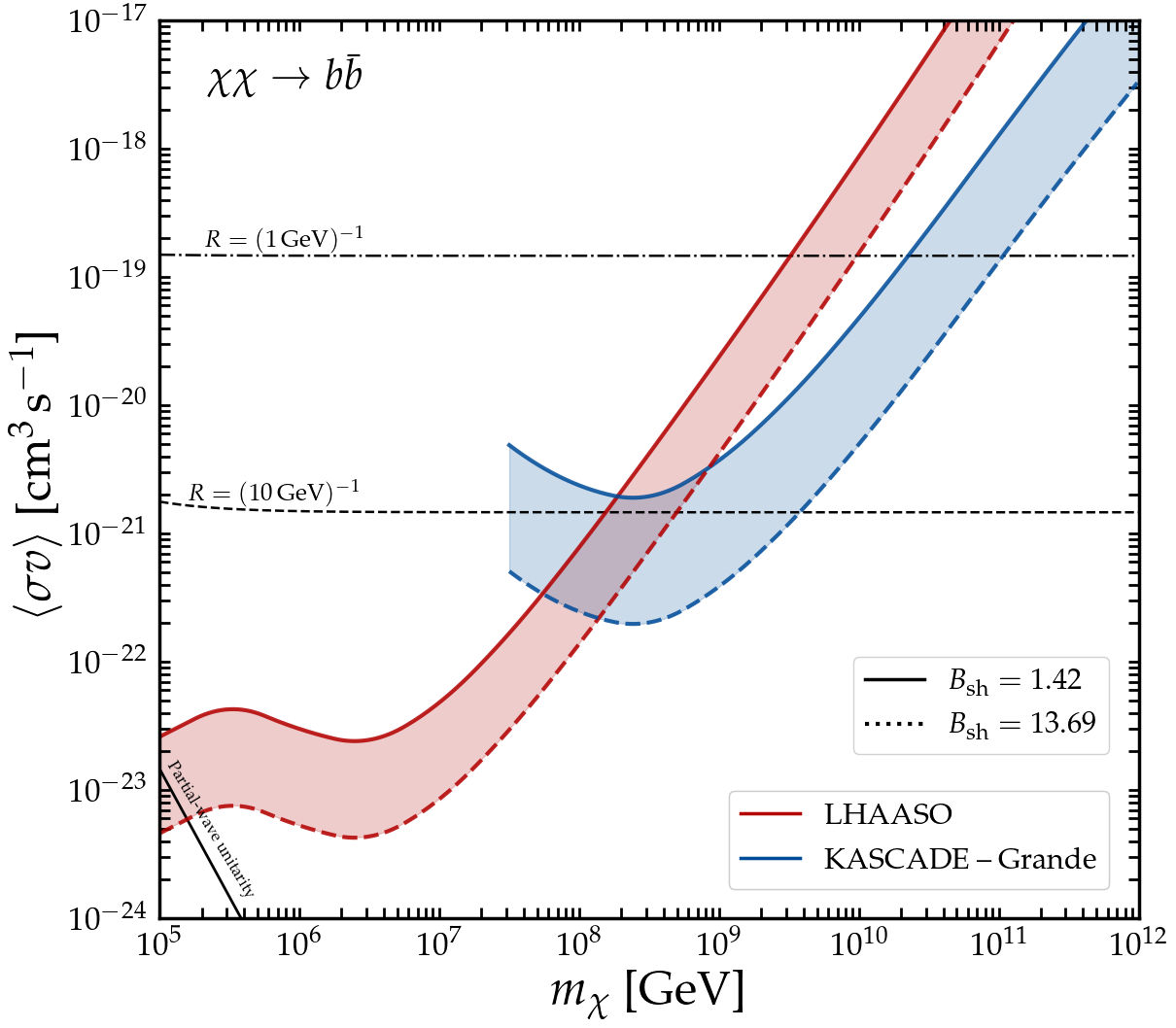}~~\\	
		\caption{Upper limits on the dark matter annihilation cross-section $\langle \sigma v \rangle$ for the $\chi \chi \to b\bar{b}$ channel. \textbf{Left panel:} The impact of background model uncertainties, illustrating the Min (solid lines) and Max (dashed lines) constraints for LHAASO and Tibet AS$_\gamma$. \textbf{Right panel:} The effect of subhalo boost factors on the exclusion limits, comparing $B_{\rm sh} = 1.42$ (solid lines) and $B_{\rm sh} = 13.69$ (dotted lines) for LHAASO and KASCADE-Grande computed based on the concentration parameters from \cite{Correa:2015} and \cite{Okoli:2016}, respectively.}
		\label{fig:background_and_boost}
	\end{center}	
\end{figure*}

We use the fitting formula from Appendix A3 of A19, which 
parametrizes $B_\text{sh}$ as a combination of two sigmoid 
functions in the host mass:
\begin{equation}
\begin{split}
    \log_{10} B_\text{sh} = \,
    &\frac{X(z)}{1 + e^{-a(z)\left(\log_{10} M_\text{host} 
    - m_1(z)\right)}} \\
    &+ c(z)\left(1 + \frac{Y(z)}{1 + e^{-b(z)
    \left(\log_{10} M_\text{host} - m_2(z)\right)}}\right),
\end{split}
    \label{eq:bsh_fit}
\end{equation}
where the redshift-dependent coefficients $X(z)$, $Y(z)$, $a(z)$, 
$b(z)$, $c(z)$, $m_1(z)$, $m_2(z)$ are tabulated in A19 for the 
Correa et al.\ (2015)~\cite{Correa:2015} concentration-mass relation, 
which we adopt as our fiducial model. For comparison, we also 
show results using the Okoli \& Afshordi (2016)~\cite{Okoli:2016} 
concentration-mass relation, for which the corresponding fitting 
coefficients are tabulated in Appendix A3 of A19. The difference in $B_{\rm sh}$ between the two models (See Figure \ref{fig:background_and_boost}) follows from their conflicting extrapolations of the $c–M$ relation below the simulation floor at $M \lesssim 10^4\,M_\odot$, where Okoli \& Afshordi \cite{Okoli:2016} predict concentrations 2–3× higher than Correa et al. \cite{Correa:2015}., and their impact on $B_\text{sh}$ is 
discussed in Section~\ref{results}. At $z = 0$, the Correa 
et al.\,\cite{Correa:2015} coefficients reduce to
\begin{align}
    &X = 2.85,\quad Y = 0.96,\quad a = 0.195,\quad b = -0.83, 
    \notag\\
    &c = -0.6,\quad m_1 = 17.4,\quad m_2 = -3.
\end{align}
For the Milky Way halo mass $M_{200} = 1.30 \times 10^{12}\,M_\odot$ \cite{McMillan2016,Cautunetal2020,Rocheetal2024},
 evaluating Eq.~\eqref{eq:bsh_fit} 
gives
\begin{equation}
    B_\text{sh} \approx 1.42.
    \label{eq:bsh_value}
\end{equation}
The effective J-factor entering our analysis is therefore
\begin{equation}
    J_\text{eff}(\hat{n}) = (1 + B_\text{sh})
    \int_\text{los} ds\; \rho^2_\chi(r(s,\hat{n})),
    \label{eq:jeff}
\end{equation}
with $(1 + B_\text{sh}) \approx 2.42$, applied uniformly over 
both the inner Galactic plane 
($15^\circ < \ell < 125^\circ$, $|b| < 5^\circ$) and 
the high-latitude ($|b| > 20^\circ$) regions.

A more precise treatment would employ a radially varying $B(r)$, 
accounting for the anti-biased spatial distribution of subhalo 
centres due to tidal disruption, the position-dependent 
concentration enhancement of Moliné et al.\cite{Moline:2016pbm}, 
and baryonic suppression of subhalos in the inner Galaxy from 
the Galactic disk and bulge~\cite{Kelley:2019,Errani:2017,Chua:2017}. 
These ingredients have not been combined self-consistently over 
the full subhalo mass range relevant here, and as A19 explicitly 
note, baryonic effects on boost factors remain largely unexplored. 
The volume-integrated $B_\text{sh}$ from H18 is directly calibrated 
against $N$-body subhalo mass functions and avoids the additional 
modeling assumptions that a radial treatment would require; we 
therefore adopt it here and leave a spatially resolved boost 
treatment to future work. Since $\langle \sigma v \rangle \propto 
(1 + B_\text{sh})^{-1}$, varying $B_\text{sh}$ across the range 
of current semi-analytic predictions \cite{Ando:2019} 
shifts our constraints by less than a factor of few, 
which we include as a systematic uncertainty on our limits.

\subsection{Extragalactic gamma ray flux}

In the extragalactic space, high energy photons can interact with low energy background photons like extragalactic background light (EBL) and CMB. As a result of this interaction, high-energy electrons and positrons are generated which can then upscatter the low-energy photons to produce high-energy photons with energy $\sim\gamma^2E_e$. This whole process is called an electromagnetic cascade. In addition, prompt electrons can directly convert low-energy photons to higher energies through ICS. As electrons lose energy on much shorter length scales than cosmological scales, one often uses the ``on-the-spot" approximation. This means the electrons and positrons are assumed to lose energy so fast that one can entirely forget about their propagation while calculating the resulting gamma-ray flux.

 We use $\gamma-\rm Cascade V4$\footnote{\textcolor{purple}{https://github.com/GammaCascade/GCascade}} to take account the electromagnetic cascade and calculate the gamma ray flux. We refer the reader to Appendix~\ref{Intro to cascade} and \ref{using gamma cascade}, as well as Refs.\,\cite{Capanema:2024nwe,Blanco:2018bbf} for more details.

The extragalactic DM annihilation signal is further enhanced by 
collapsed structures along the line of sight. We account for this 
through the cosmological clumping factor $Z_c(z)$, defined as 
$\langle\rho^2_{\chi}(z)\rangle/\bar\rho^2_{\chi}(z)$, which 
boosts the extragalactic flux above the smooth-universe prediction, where $\langle\rho^2_{\chi}(z)\rangle$ denotes average of the DM density squared and $\bar\rho^2_{\chi}(z)$ denotes the average DM density sqaured, both at the redshift of $z$. The computation of $Z_c(z)$, built from the linear power spectrum, 
the Sheth--Tormen halo mass function, and NFW profiles with 
substructure, is described in detail in 
Appendix~\ref{app:clumping}.
 
 We add all of the above contributions to get the total gamma-ray flux from DM annihilation

\begin{equation}
\frac{d^2\Phi_{\mathrm{tot}}}{dE_\gamma \, d\Omega}
=
\frac{d^2\Phi^{\mathrm{G}}}{dE_\gamma \, d\Omega}
+
\frac{d^2\Phi^{\mathrm{G}}_{\mathrm{ICS}}}{dE_\gamma \, d\Omega}
+
\frac{d^2\Phi^{\mathrm{EG}}}{dE_\gamma \, d\Omega}\,\,,
\end{equation}
where the superscript G denotes the Galactic component and EG denotes the extragalactic components. 

For the DM mass range considered, the Galactic prompt flux dominates across most energies relevant to the detectors used in this work; the Galactic ICS sits at $\mathcal{O}(10\%)$, with the extragalactic component a further order of magnitude below.

\section{Data \& Analysis}
We use diffuse gamma-ray datasets from Tibet AS$_\gamma$ and LHASSO, and UHE gamma-ray upper limits from Tibet AS$_\gamma$, KASCADE, KASCADE-Grande, PAO, and TA detectors. Each of these detectors have specific energy ranges of interest and sensitivity. In the following, we discuss the datasets and upper limits of interest. We also outline the analysis methods we use to obtain the limits on heavy annihilating DM.
\label{dataset}

\subsection*{\textit{Datasets}}
\subsection{Tibet AS$_\gamma$
\,\cite{TibetASgamma:2021tpz}}

Tibet AS$_\gamma$ is located at Yangbajing, Tibet. The experiment consists of surface air-shower array combined with underground
water-Cherenkov muon detectors. The collaboration was the first to detect sub-PeV diffuse gamma ray emission in the energy range from 100 TeV to 1 PeV \cite{TibetASgamma:2021tpz}. These measurements are consistent with various astrophysical models where primarily Galactic cosmic rays interact with interstellar gas\,\cite{Fang:2021ylv, Dzhatdoev:2021xjh,Qiao:2021iua,Liu:2021lxk, Koldobskiy:2021cxt, DeLaTorreLuque:2025zsv}. The measurements by Tibet AS$_\gamma$ are reported for two different regions (i) $25^\circ < l < 100^\circ$, $|b|<5^\circ$ and (ii) $50^\circ<l<200^\circ$, $|b|<5^\circ$. 

\subsection{LHAASO\,\cite{LHAASO:2023gne, LHAASO:2024lnz}}
The Large High Altitude Air
Shower Observatory (LHAASO), situated in China, has two main detectors for detecting high energy diffuse gamma rays -- the Water Cherenkov Detector Array (WCDA) and the Square Kilometer Array (KM2A).
LHAASO collaboration has made  measurements of diffuse gamma ray from the Galactic plane with both its KM2A and WCDA arrays with energies from 1 TeV to 1000 TeV\,\cite{LHAASO:2024lnz, LHAASO:2023gne} . The measurement is divided into (i) inner Galactic plane ($15^\circ<l<125^\circ$, $|b|<5^\circ$) and (ii) outer Galactic plane ($125^\circ<l<235^\circ$, $|b|<5^\circ$) regions. The LHAASO dataset used in this work is available in the supplementary material of Ref.\,\cite{LHAASO:2024lnz}. We note that, though the LHAASO collaboration noted excess flux of gamma rays over their background model, subsequent modelings of the background taking into account the various uncertainties and improved masking have shown background models to be consistent with the measurements from LHAASO\,\cite{Chen:2024yin,Zhang:2023ajh,He:2025oys,Ambrosone:2025wxc,Castro:2025wgf,Kato:2025gva,DeLaTorreLuque:2025zsv, He:2025oys, Prevotat:2025ktr,Vecchiotti:2024kkz,Kaci:2024lwx,Kaci:2024wra,Dekker:2023six,Shao:2023aoi}.

\subsection*{\textit{Background Model}}

For the Galactic plane diffuse gamma-ray datasets from Tibet~AS$\gamma$ and 
LHAASO, an accurate model of the astrophysical background is required for our  analysis. We adopt the $\gamma$-optimized diffuse 
emission models of Ref.~\cite{DeLaTorreLuque:2025zsv} as our 
background templates. These models are computed numerically using 
the \texttt{DRAGON2} and \texttt{HERMES} codes, with cosmic ray injection spectra tuned 
to locally measured proton and helium data, and account for $\gamma$\,--\,$\gamma$ 
opacity from CMB scattering.

To bracket the uncertainty in the cosmic ray spectral shape at $E \gtrsim 100$~TeV, 
Ref.~\cite{DeLaTorreLuque:2025zsv} consider two configurations: the \textit{Min} setup, tuned to KASCADE\,\cite{Apel:2013uni}, 
CALET\,\cite{CALET:2019bmh}, and DAMPE\,\cite{DAMPE:2019gys} measurements, and the \textit{Max} setup, tuned to the harder 
IceTop spectrum\,\cite{IceCube:2019hmk}. As we will show later, the Max background model provides a stronger constraint compared to the Min model. Thus we adopt the Min model as our 
benchmark to derive conservative limits on heavy annihilating DM. We note that the Max setup is 
independently disfavored by LHAASO data at ${>}\,10\sigma$~\cite{DeLaTorreLuque:2025zsv}, 
further motivating Min as the physically preferred background.

\subsection*{\textit{Analysis}}
For the above gamma-ray datasets, we use a $\chi^2$ estimator to set limits of the annihilation cross-section $\langle\sigma v\rangle$ for a given DM mass, where $\chi^2$ is given by\,\cite{Cowan:2010js, Lamperstorfer:2015cfg} 
\begin{equation}
    \chi^2=\sum_i\frac{(\phi_{\rm data}^i-\phi_{\rm model}^i-\phi_{\rm \chi}^i)^2}{\sigma_i^2}\,.
\end{equation}
Here $\phi_{\rm data}^i$ is the measured differential gamma ray flux, $\phi_{\rm model}^i$ is the differential astrophysical gamma ray flux, $\phi_{\rm \chi}^i$ is the differential flux from DM annihilation, and $\sigma_i$ is the uncertainty in differential flux measurement in the $i^{\rm th}$ energy bin. We set the limit by first finding the cross-section that minimizes $\chi^2$ for a given mass, and then cross-sections for which $\chi^2-\chi_{\rm min}^2\geq 2.71$ are excluded for a given mass at $95\%$ C.L.
\subsection*{\textit{Upper Limits}}
\subsection{Tibet AS$_\gamma$
\,\cite{Neronov:2021ezg}} 
Ref.\,\cite{Neronov:2021ezg} used the all sky cosmic ray measurement by Tibet AS$_\gamma$, to set upper limits on the diffuse gamma ray flux in the high Galactic latitudes. This was performed for $|b|>20^\circ$ in the energy range 100 TeV to 1 PeV. The authors used the all sky cosmic ray data and applied the appropriate muon cut. Assuming the remaining muon-poor events are all coming from gamma-ray induced shower, they obtained upper limits on diffuse gamma-ray flux away from the Galactic plane.  
\subsection{KASCADE-Grande\,\cite{2017ApJ...848....1A}} The KArlsruhe Shower Core and Array DEtector (KASCADE), located in Germany, consisted of three detector systems-- a field array, a muon tracking detector, and a central detector. Later, KASCADE-Grande was built as an extension to the KASCADE array. Using their datasets, KASCADE and KASCADE-Grande have set upper-limits on the diffuse gamma ray flux from 100 TeV to 1 EeV. This analysis was based on identifying showers with low-muon count as expected for gamma-ray induced showers. We use the available integrated flux upper limits from Ref.\,\cite{2017ApJ...848....1A}. The dataset is shown by purple triangles in Figure~\ref{fig:flux_plots}.

\subsection{PAO\,\cite{PierreAuger:2015fol, PierreAuger:2016kuz}}
Pierre Auger Observatory (PAO), located in Argentina, is measuring the cosmic ray spectrum beyond 100 EeV. So far, there has been no confirmed photon detection, but PAO has upper limits on the integrated diffuse flux using its (i) Surface Detectors (SD) and (ii) Hybrid Detectors (HD). SD measurements go to much higher energies and have more statistics, while HD has superior energy resolution. We use the upper limits from Ref.\,\cite{PierreAuger:2022aty} obtained using SD, and from Ref.\,\cite{PierreAuger:2016kuz} using HD, denoted by dark green squares and light pink triangles in Figure~\ref{fig:flux_plots}, respectively. Throughout this work, we will refer to the the SD measurements by PAOSD and HD measurements by PAOhy.
\subsection{TA\,\cite{TelescopeArray:2018rbt}} 
The Telescope Array (TA), located in USA, is another air shower detector that uses a combination of surface detectors and fluorescence detectors. TA has obtained upper limits on UHE diffuse gamma rays above $10^{18}$ eV using nine years of dataset from their surface detector array\,\cite{TelescopeArray:2018rbt}. These upper limits have been obtained by comparing both simulation results and datasets for various air-shower parameters like curvature, width of the shower font, steepness of the lateral distribution function, and timing. The upper limit is shown by orange diamonds in Figure~\ref{fig:flux_plots}. 
\subsection*{\textit{Analysis}}

In case of upper limits, a cross-section for a particular DM mass, $m_\chi$, is excluded if the photon flux from DM annihilation exceeds the flux upper limits, $\Phi_{\rm \chi}^i> \Phi_{\rm UL}^i$, for any of the $i^{th}$ energy bins. As upper limits are reported in terms of the differential flux of photons integrated above energy $E_0$, we calculate the differential fluxes integrated above $E_0$ for a given mass $m_{\chi}$ and cross-section $\langle\sigma v\rangle$  and then set the limits. We note that given an upper limit is not a detection, we refrain from using any particular astrophysical background models in this case. Inclusion of any background model will strengthen our limits. Therefore our approach and the resulting limits are conservative.
\begin{figure}[t]
\centering
\includegraphics[width=\columnwidth]{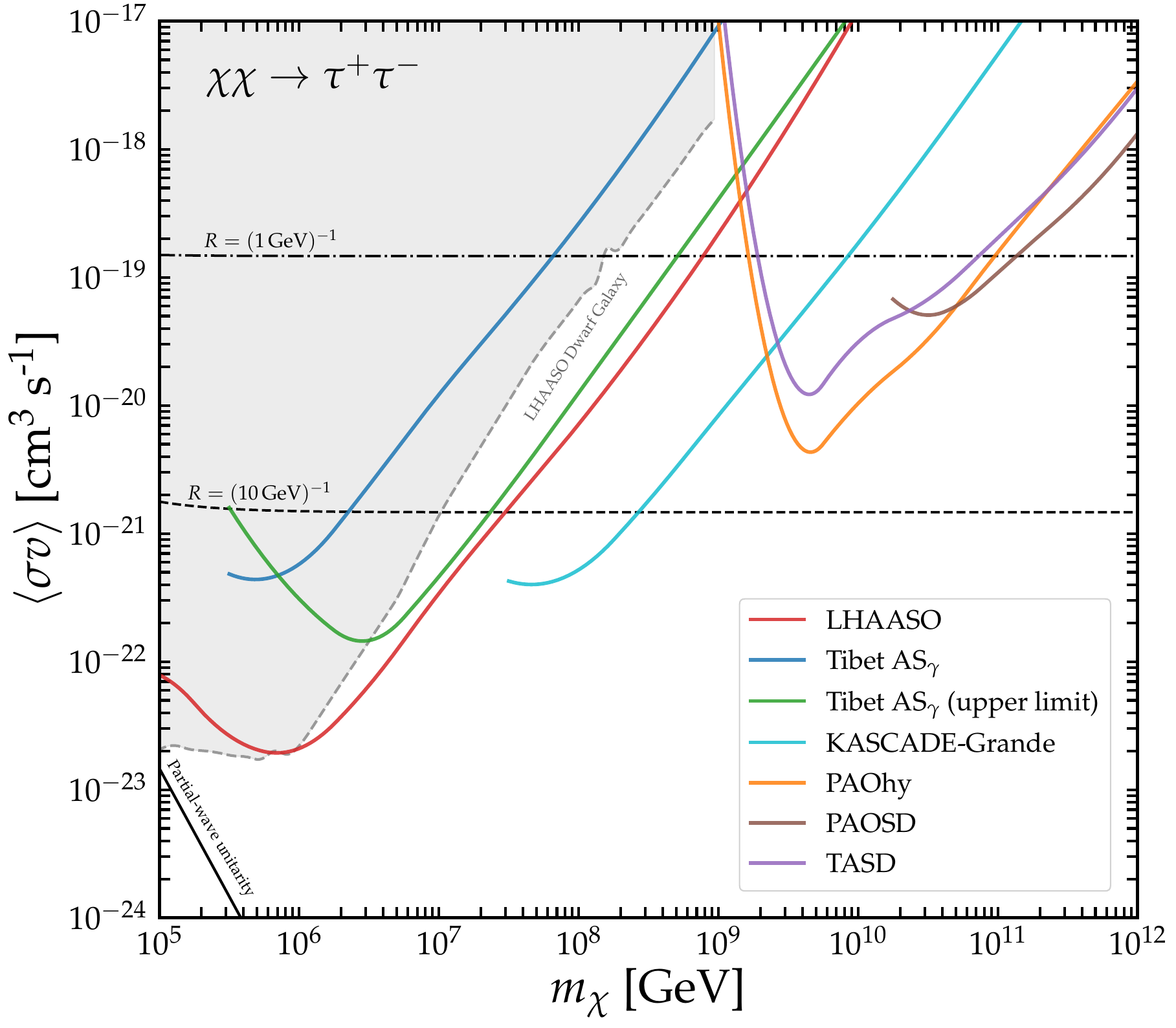}
\caption{Constraints on DM annihilation cross section, $\langle\sigma v\rangle$ as a function of DM mass $m_\chi$ for the annihilation channel $\chi\chi\to \tau^{+}\tau^{-}$ (100\% branching ratio) using different gamma-ray datasets. All the existing limits and benchmarks are same as Figure \ref{fig:bbbar_constraints}. }
\label{fig:ee_constraints}
\end{figure}

\section{Results and discussion}
\label{results}

In Figure~\ref{fig:flux_plots}, we show the differential gamma-ray flux integrated above the energy $E_0$ for DM annihilation. We take three benchmark DM masses $\rm 10^8\,GeV$, $\rm 4\times10^9\,GeV$, and $\rm 10^{12}\,GeV$ with annihilation cross-section of $\rm 1.8\times10^{-19}\,cm^3s^{-1}$, considering only prompt photon emission (prompt electron and EG contribution being negligible). The left and right panels of  Figure~\ref{fig:flux_plots} show the gamma-ray fluxes for DM annihilation channels $\chi\chi\to b\bar{b}$ and $\chi\chi\to \tau^{+}\tau^{-}$, respectively. Along with the DM-originated gamma-ray flux we show the upper limits on diffuse gamma-ray from KASCADE-Grande, PAO, and TA. Comparing the expected gamma-ray flux with the upper limits, one can obtain constraints on DM masses and annihilation cross sections. For example, in case of $\chi\chi\to b\bar{b}$ (left panel of Figure~\ref{fig:flux_plots}) for DM masses $\rm 10^8\,GeV$ and $\rm 4\times10^9\,GeV$ with annihilation cross section $\rm 1.8\times10^{-19}\,cm^3s^{-1}$, the expected gamma-ray flux overshoots the upper limits from KASCADE-Grande at multiple energy bins. Therefore the respective masses and cross-sections are ruled out by observations from KASCADE-Grande. For flux comparison with $\rm Tibet\,AS_\gamma$ and LHAASO datasets, we direct the readers to Refs.\,\cite{Boehm:2025qro, Dubey:2025ouh, Maity:2021umk}.

In Figure~\ref{fig:bbbar_constraints} and Figure~\ref{fig:ee_constraints}, we show our constraints for DM annihilating to $b\bar{b}$ and $\tau^+\tau^-$ channels, respectively. For these limits we use the gamma-ray measurements from LHAASO, $\rm Tibet\,AS_\gamma$ along with the upper limits from $\rm Tibet\,AS_\gamma$, KASCADE-Grande, PAO, and TA. The  LHAASO, $\rm Tibet\,AS\gamma$ limts are evaluated taking into account the Min background model. Harder gamma-ray spectra produced for the $\tau^+\tau^-$ channel lead to stronger constraints than the $b\bar{b}$ channel. Our work improves upon existing limits\,\cite{LHAASO:2024upb,VERITAS:2024usn,Acharyya:2023ptu} in the parameter space significantly.  We also extend the limits to larger masses, which have remained unexplored so far. This extension become possible because of the gamma-ray upper limits from KASCADE-Grande, PAO, and TA. We show the partial-wave unitarity bound by black solid line. The benchmark unitarity limits for DM radii $R= (1 \,\rm GeV)^{-1}$ and $R= (10 \,\rm GeV)^{-1}$ are shown by the black dot-dashed and dashed lines, respectively. Parameter spaces above these lines are ruled out from unitarity arguments for composite DM (see section \ref{unitarity} for a detailed discussion). We find that our work provides stringent limits on viable heavy annihilating DM parameter space allowed by unitarity.

The relative strength of LHAASO and $\rm Tibet\,AS_\gamma$ limits in Figure~\ref{fig:bbbar_constraints} and Figure~\ref{fig:ee_constraints} can be understood by comparing the corresponding measurements along with the astrophysical background model. In Figure 5 and 6 of Ref.\,\cite{DeLaTorreLuque:2025zsv}, for the Min background model, we see that the LHAASO measurements are fitted well in a wide energy range whereas $\rm Tibet\,AS_\gamma$ measurements are above the background predictions. As a result for the Min model, the constraints on DM annihilation cross sections from LHAASO datasets are stronger compared to $\rm Tibet\,AS_\gamma$, evident in Figure~\ref{fig:bbbar_constraints} and Figure~\ref{fig:ee_constraints} in our work. The limits derived from UHE gamma-ray upper limits can also be understood similarly. The limits derived from KASCADE-Grande, PAO, and TA start at different DM masses. This is because the gamma-ray spectra endpoint is determined by the DM mass. Only when the DM originated gamma-ray flux overlaps with at least one observation bin, we can constrain the corresponding DM parameters. The relative strength in these limits is dictated by the strength of the corresponding upper limits. For example, the upper limits from PAOhy are stronger compared to TA. As a result, the PAOhy limits on DM annihilation are stronger than the TA limits, as seen in Figure~\ref{fig:bbbar_constraints} and Figure~\ref{fig:ee_constraints}.  

The constraints coming from LHAASO and $\rm Tibet\,AS_\gamma$, in Figure~\ref{fig:bbbar_constraints} and Figure~\ref{fig:ee_constraints} are obtained assuming the Min background model with a subhalo boost factor $\rm B_{sh}$ of 1.42. We show the effect of the background model as well as the Galactic subhalo boost factor on our constraints in Figure~\ref{fig:background_and_boost}. As shown in the left panel, the Max model leads to stronger constraints than Min model. Therefore, the constraints we show are conservative. The larger boost factor also leads to stronger constraints, as shown in the right panel of Figure~\ref{fig:background_and_boost}. Again, we find that our results in Figure~\ref{fig:bbbar_constraints} and Figure~\ref{fig:ee_constraints} are conservative. 

Both LHAASO and $\rm Tibet\,AS_\gamma$ have measured diffuse gamma rays from the inner and outer regions of the Galactic plane region. The limits obtained in Figure~\ref{fig:bbbar_constraints} and Figure~\ref{fig:ee_constraints} are dominated by inner region datasets from both of the experiments. This is because the gamma-ray flux from annihilation scales as $\rho_\chi^2$. Hence, the inner region contributes significantly to DM annihilation compared to the outer region of the Galactic plane. 

The limits presented in our work do not change appreciably with DM density profile. This is because none of the datasets used here observe the Galactic centre region, where the possible DM profile variations are maximum. We find that depending on the DM density profile, our constraints change at most by a few percent. Given the much larger uncertainties from the underlying background models, our benchmark NFW results are robust.

In our work, we include the effects of SL, IR, and CMB backgrounds on high-energy gamma rays. For photon energies $\geq$ a few times $10^{18}\rm\,eV$, Cosmic Radio background (CRB) affects the propagation of gamma rays\,\cite{Nitu:2020vzn}.  Extragalactic DM originated gamma-ray flux is small, and the mean free path for energies $\geq\rm few\, times\,10^{18}\,eV$ is $\gtrsim\,\rm Mpc$, making its effect entirely negligible for Galactic gamma-ray flux. Thus, in our work we do not take into account the effects of CRB.  

Angular datasets at these high energies can additionally help in discriminating DM contributions from astrophysical backgrounds. We performed such an analysis in Ref.\,\cite{Dubey:2025ouh}. We expect the limits to be competitive with the total flux constraints. Therefore, we leave out a detailed morphological study for future analysis.\\

In summary, we show that observations by high energy air-shower detectors can provide stringent constraints on heavy annihilating DM beyond the simplest unitarity considerations. Upcoming observations of high energy gamma rays by various telescopes can thus unveil the nature of DM.\\

\textit{Note added:} While this work was in progress, Ref.\,\cite{Boehm:2025qro} did a similar analysis where they used LHAASO measurements to put limits on heavy annihilating DM.

\section*{Acknowledgements} 
We especially thank Ranjan Laha and Tarak Nath Maity for detailed discussions.
We thank Ranjan Laha for helpful comments on the
manuscript. We also thank Debajit Bose, Subhadip Bouri, Marco Chianese, Anirban Das, Stefano Morisi, Kenny C.\,Y.\,Ng, and Abhijeet Singh for useful comments and discussions.  A.D.
acknowledges the financial support provided by the Ministry of Education (MoE), Government of India. D.J.D.
acknowledges the financial support provided by the Ministry of Education (MoE), Government of India.
A.K.S. acknowledges the Ministry of Human Resource Development, Government of India, for financial support via the Prime Minister’s Research Fellowship (PMRF).

\bibliographystyle{JHEP}
\bibliography{ref.bib}

\clearpage
\onecolumngrid
\thispagestyle{plain}

\appendix

\section{Brief description of  $\gamma-{\tt Cascade}\, V4$}
\label{Intro to cascade}

High-energy gamma rays undergo absorption while travelling over cosmological distances. They interact with low-energy photons, mainly Cosmic Microwave Background (CMB) and Extragalactic Background Light (EBL), to produce energetic $e^{+}e^{-}$ pairs. These electrons and positrons can further scatter low-energy photons from the CMB or EBL to high-energy gamma rays. This way, an avalanche of photons, electrons, and positrons is produced. This process is called \textbf{Electromagnetic Cascade}.

Mathematically, we want to know the following:\,\,given an initial high-energy photon source luminosity or a distribution of sources with some photon luminosities, what is the final flux that we get? 

$\gamma$-{\tt Cascade} tries to answer this question by breaking the propagation into a set of redshift bins and iteratively solving the cascade equations. We note that Refs.\,\cite{Blanco:2018bbf, Capanema:2024nwe}
provide a more detailed discussion about the setup. We discuss the main ingredients here for completeness. 

 Let's first discuss the prompt photon spectrum. We propagate the prompt photons for a distance corresponding to a redshift bin of $\delta\,z=10^{-6}$. Some of the photons will be absorbed by CMB/EBL photons to produce $e^{-}e^+$ pairs. In the next step, we calculate the photon spectra emitted by these electrons and positrons. Given that these electrons and positrons lose energy very fast, we simply add the photons produced by the electrons and positrons to the unabsorbed photons.  Having done this, we again let the photons propagate and repeat the same set of steps as above until the photons reach us. Note that, while propagation is happening, we inject prompt photons from the dark matter annihilation after every redshift traversal of redshift $\Delta z$ ($\Delta z$ is defined more clearly below).
 
 For Prompt electron spectra, we use the “on-the-spot-ics-spec” photon spectrum, produced by electrons. Here we note that the validity of “on-the-spot-ics-spec” has been shown in Ref.\,\cite{Capanema:2024nwe} by comparing it with Monte Carlo-based code {\tt CRbeam}, which in turn is shown to be consistent with {\tt ELMAG} in Ref.\,\cite{Kalashev:2022cja}. Initially, there are no photons, but they are produced as soon as the electrons are injected, given their extremely short energy loss length scale. Once we have photons, we proceed with the same set of steps that we followed for photons, modulo an injection of electrons from a particular dark matter annihilation.\\

Let us introduce a few key ingredients that go into $\gamma$-{\tt Cascade} V4 calculation in the context of DM annihilation.

\subsection{Diffuse steps}

\textbf{Diffuse steps} is a way of binning the redshift at which prompt photons (and electrons) are injected. We show them in the following table

\begin{table}[H]
\centering
\large
\begin{tabular}{|c|c|c|c|c|}
\hline
{Phase} & {Steps} & {$\Delta z$} & {$z$ range} & {Index} \\
\hline
1 & 10  & $10^{-6}$ & $0 \to 10^{-5}$         & 1--10    \\[4pt]
2 & 9   & $10^{-5}$ & $10^{-5} \to 10^{-4}$   & 11--19   \\[4pt]
3 & 9   & $10^{-4}$ & $10^{-4} \to 10^{-3}$   & 20--28   \\[4pt]
4 & 9   & $10^{-3}$ & $10^{-3} \to 10^{-2}$   & 29--37   \\[4pt]
5 & 999 & $10^{-2}$ & $10^{-2} \to 10$        & 38--1036 \\[4pt]
\hline
\end{tabular}
\caption{Redshift binning: 1036 steps covering $z = 0$ to $10$.}
\label{tab:redshift_bins}
\end{table}

Each of these steps is further divided into steps of $10^{-6}$ (no such division required for the first 10 steps), on which \textbf{CascadeCycle} operates.

\subsection{CascadeCycle}

\textbf{CascadeCycle} operates on very small steps of $\delta z = 10^{-6}$ in redshift. 
It takes an initial photon spectrum $\frac{dN}{dE_{\gamma}}(E_{\gamma})$. 
If the optical depth at energy $E_{\gamma}$ is ${\tau_{\mathrm{PP}}(E_{\gamma}, \delta z)}$, then
\[
\frac{dN}{dE_{\gamma}}(E_{\gamma}) \, e^{-\tau_{\mathrm{PP}}(E_{\gamma}, \delta z)},
\]
do not undergo absorption, and the part of the spectrum undergoing absorption is given by
\[
\frac{dN}{dE_{\gamma}}(E_{\gamma}) 
\left(1 - e^{-\tau_{\mathrm{PP}}(E_{\gamma}, \delta z)}\right).
\]

$\gamma$-{\tt Cascade} uses the ``on-the-spot approximation'' to account for the photon spectrum produced from the electrons and positrons produced from the absorbed high-energy photons. It assumes that all the produced electrons and positrons practically lose all of their energy while traversing $\delta z = 10^{-6}$. Therefore, the secondary photon spectrum produced due to ICS after traversing a redshift bin of width $\delta z$ is given by

\[
\frac{dN_{\gamma,\mathrm{OTS}}}{dE'_{\gamma}}(E'_{\gamma}, z)
=
\int dE_{\gamma} \,
\left\{
\frac{dN_{\gamma \to e \to \gamma,\mathrm{OTS}}}{dE'_{\gamma}}
(E'_{\gamma}, E_{\gamma}, z)
\times
\left[
\left(1 - e^{-\tau_{\mathrm{PP}}(E_{\gamma}, \delta z)}\right)
\frac{dN_{\gamma}}{dE_{\gamma}}(E_{\gamma}, z)
\right]
\right\}
\]

The total spectrum is the attenuated spectrum plus the photons produced ``on-the-spot'' from the electrons and positrons:
\[
\frac{dN_{\gamma,\mathrm{tot}}}{dE'_{\gamma}}(E'_{\gamma}, z)
=
\left[
e^{-\tau_{\mathrm{PP}}(E_{\gamma}, \delta z)}
\frac{dN_{\gamma}}{dE_{\gamma}}(E_{\gamma}, z)
\right]_{E_{\gamma} = E'_{\gamma}}
+
\frac{dN_{\gamma,\mathrm{OTS}}}{dE'_{\gamma}}(E'_{\gamma}, z)
\]

The spectrum of photons produced by the electrons and positrons is stored in the form of precalculated tables, which can be found in the ``cycle-spec'' directory.\\

The previous calculations, requires the optical depths $\tau_{\mathrm{PP}}(E_{\gamma}, \delta z)$, which can be calculated using
\[
\tau_{\mathrm{PP}}(E_{\gamma})
=
\frac{\ell}{\lambda_{\mathrm{PP}}(E_{\gamma}, z^\ast)},
\]
with
\[
\ell
=
\int_{z_i}^{z_{i+1}}
\frac{c \, dz}{(1+z) H(z)},
\]
corresponding to the redshift interval
\[
\delta z = z_{i+1} - z_i = 10^{-6},
\]
where $z^\ast$ is the redshift where the cascade is happening. For calculations, it is set to the value of the nearest redshift found in \textbf{zReg} array.

\subsection{zReg}

It is a uniform redshift binning of redshift in steps of $\Delta z = 0.01$. It is an array of size 1001. Various precalculated tables are calculated at the values of \textbf{zReg}.

\subsection{energies}

It is an array of size 300. It is a list of energies from $10^{-1}$ to $10^{12}$ GeV with uniform logarithmic bins.

For a more comprehensive discussion on various subtle points, including various definitions, we refer the reader to \cite{Capanema:2024nwe}

\section{Using $\gamma$-{\tt Cascade} V4 for EG DM annihilation}
\label{using gamma cascade}
For any two-body annihilation channel, let's say for $b\bar{b}$, we would like to know how many photons we would get in a detector. But $b\bar{b}$ going to electrons will also give photons through inverse Compton scattering. We add these two contributions to get the total photons from the $b\bar{b}$ channel.

The quantity that should be injected is the differential rate of production of photons per unit volume, which is given by 
\begin{equation}
\begin{aligned}
\frac{1}{2} n_{\rm DM}^2 \langle\sigma v\rangle 
\frac{dN}{dE_\gamma}(E_\gamma)
&\to
\underbrace{\frac{1}{2}\langle\sigma v\rangle\left(\frac{\rho_c \Omega_{\rm \chi}}{m_{\rm \chi}}\right) 
\frac{dN}{dE_\gamma}(E_\gamma)}_{\text{annihilation rate to photons}}
\,
\underbrace{\left(\frac{\rho_c \Omega_{\rm \chi}}{m_{\rm \chi}}\right)
(1+z)^6 Z_c(z)}_{\text{distribution of sources}}\,\,.
\end{aligned}
\label{eq:injection}
\end{equation}

We can now call \textbf{CascadeDiffuse} with \textbf{CascadeDiffuse} [annihilation rate to photons, $Z_{\rm max}$, distribution of sources], where $Z_{\rm max}$ is chosen in a way that leads to an approximate saturation of the final photon spectrum. On the right, we also include the clumping factor, $Z_c(z)$.
This can be straight forwardly implimented for the channels directly giving us the prompt photon flux. 

For the second case, as we discussed, dark matter annihilation channels can give rise to an $e^+e^-$ prompt spectrum, which can undergo inverse Compton scattering to produce photons. For this, we use the pre-computed “on-the-spot-ics-spec” (OTSICS) and integrate over the electron energies to obtain the ICS photon spectrum
\begin{equation}
\frac{dN_{\gamma,\mathrm{OTSICS}}}{dE'_{\gamma}}(E'_{\gamma}, z)
=
\int dE_{e}\,
\left\{
\frac{dN_{ e \to \gamma,\mathrm{OTSICS}}}{dE'_{\gamma}}
(E'_{\gamma}, E_{e}, z)
\times
\frac{dN_{e}}{dE_{e}}(E_{e}, z)
\right\}\,\,.
\label{eq:otsics}
\end{equation}
This ICS photon spectrum is added in the \textbf{very first step} of the \textbf{CascadeCycle}.

To handle both of these in one shot, we enable \textbf{CascadeDiffuse} to handle four arguments-
\textbf{CascadeDiffuse} 
 [annihilation rate to photons, annihilation rate to electrons and positrons, $Z_{\rm max}$, distribution of sources]. The photon spectrum is handled as usual, according to equation \eqref{eq:injection}. For the electrons and positrons, we use equation \eqref{eq:otsics} to calculate the photon spectrum, and this is added to the otherwise photon spectrum at every \textbf{Diffuse steps}.

\section{The Extragalactic Clumping Factor $Z_c(z)$}
\label{app:clumping}

The source term in Eq.~\eqref{eq:injection} of Appendix~\ref{using gamma cascade}
contains the factor $Z_c(z) \equiv \langle\rho^2_{\chi}(z)\rangle
/ \bar\rho^2_{\chi}(z)$, the ratio of the volume-averaged squared
DM density to the smooth-universe mean squared. Since DM annihilation
scales as $\rho_\chi^2$, the cosmological signal is set by this second
moment rather than the mean. We compute $Z_c(z)$ following the
halo-model framework of~\cite{Ando:2013wia}, specifically their
Eqs.~(3) and~(6), with the power spectrum and mass variance built
as in~\cite{Eisenstein:1997jh} and the halo mass function following
the implementation of~\cite{Murray:2013qza}.

\textit{Linear matter power spectrum.}---We adopt the Planck
2018~\cite{Planck:2018vyg} cosmology. The CDM+baryon transfer
function decomposes as~\cite{Eisenstein:1997jh}
\begin{equation}
T_{cb}(k,z) \;=\; T_{\rm master}(k)\,\frac{D_{cb}(k,z)}{D_1(z)},
\label{eq:Tcb}
\end{equation}
where $T_{\rm master}(k)$ is the time-independent master function,
$D_1(z)$ is the large-scale linear growth factor, and $D_{cb}(k,z)$
is the scale-dependent CDM+baryon growth function incorporating
neutrino free-streaming suppression~\cite{Eisenstein:1997jh}.
The linear matter power spectrum is then~\cite{Eisenstein:1997jh}
\begin{equation}
P(k,z) \;=\; \delta_H^2\,\frac{2\pi^2}{k^3}
\!\left(\!\frac{c\,k}{H_0}\!\right)^{\!\!3+n_s}
T_{cb}^2(k,z)\!\left[\frac{D_1(z)}{D_1(0)}\right]^2,
\label{eq:Pkz}
\end{equation}
where $\delta_H$ is the COBE-normalized amplitude~\cite{Eisenstein:1997jh}
and $n_s = 0.965$. Note that $D_1$ rather than $D_{cb}$ enters the
growth ratio; the scale-dependent suppression is already absorbed
into $T_{cb}$.

\textit{Mass variance.}---The rms variance at Lagrangian radius $R$
is~\cite{Murray:2013qza}
\begin{equation}
\sigma^2(R,z) \;=\; \frac{1}{2\pi^2}
\int_0^\infty k^2\,P(k,z)\,W^2(kR)\,dk,
\label{eq:sigma}
\end{equation}
with the Fourier-space top-hat $W(kR) = 3[\sin(kR) -
kR\cos(kR)]/(kR)^3$~\cite{Murray:2013qza} and the
mass--radius relation $M = (4\pi\rho_0/3)R^3$ defining $R(M)$,
where $\rho_0$ is the comoving mean matter density.

\textit{Halo mass function}---The logarithmic derivative entering
the Halo Mass Function (HMF) is~\cite{Murray:2013qza}
\begin{equation}
\frac{d\ln\sigma}{d\ln M} \;=\;
\frac{3}{2\sigma^2\pi^2 R^4}
\int_0^\infty \frac{dW^2}{dM}\,\frac{P(k,z)}{k^2}\,dk,
\label{eq:dlnsig}
\end{equation}
where~\cite{Murray:2013qza}
\begin{align}
\frac{dW^2}{dM} \;=\;
&\bigl[\sin(kR) - kR\cos(kR)\bigr]
\times\;\left[\sin(kR)\!\left(1 - \frac{3}{(kR)^2}\right)
+ \frac{3\cos(kR)}{kR}\right].
\label{eq:dW2dM}
\end{align}
The integrals in Eqs.~(\ref{eq:sigma}) and~(\ref{eq:dlnsig}) are
evaluated in two pieces — $k \in [0,\,0.1/R]$ and
$k \in [3/R,\,\infty)$ — following the convergence bounds
of~\cite{Murray:2013qza}, which ensure that 95\% of the
window-function integral is captured at both ends. 

The comoving HMF reads~\cite{Murray:2013qza}
\begin{equation}
\frac{dn}{dM}(M,z) \;=\; M\cdot\frac{\rho_0}{M^2}\,f(\sigma)\,
\left|\frac{d\ln\sigma}{d\ln M}\right|,
\label{eq:HMF}
\end{equation}
with the Sheth--Tormen multiplicity function~\cite{Sheth:1999mn,
Sheth:2001dp}
\begin{equation}
f(\sigma) \;=\; A\sqrt{\frac{2a}{\pi}}
\left[1 + \left(\frac{\sigma^2}{a\delta_c^2}\right)^{\!p}\right]
\frac{\delta_c}{\sigma}\,
\exp\!\left(-\frac{a\delta_c^2}{2\sigma^2}\right),
\label{eq:fsigma}
\end{equation}
where $\delta_c = 1.686$~\cite{Press:1974} and
$(A,\,a,\,p) = (0.3222,\,0.707,\,0.3)$. We have verified that
the resulting $dn/d\log M$ is consistent with the tabulated
{\tt HMFcalc} output of~\cite{Murray:2013qza} over
$10^8 \lesssim M/M_\odot \lesssim 10^{14}$.

\textit{Single-halo squared-density integral.}---Each host of
virial mass $M$ is assigned an NFW profile~\cite{Navarro:1996gj}
\begin{equation}
\rho_{\chi}(r) \;=\;
\frac{\rho_s}{(r/r_s)(1 + r/r_s)^2},
\label{eq:NFW_app}
\end{equation}
with the virial radius defined through $M = 4\pi r_{\rm vir}^3
\Delta_{\rm vir}(z)\rho_c(z)/3$, where
$\Delta_{\rm vir}(z) = 18\pi^2 + 82d - 39d^2$ and
$d = \Omega_m(1+z)^3/[\Omega_m(1+z)^3 + \Omega_\Lambda] - 1$~\cite{Ando:2013wia}.
The scale density $\rho_s$ is fixed by requiring $\int dV\,\rho_{\rm host} = M$~\cite{Ando:2013wia}.
For the concentration $c_{\rm vir} = r_{\rm vir}/r_s$ we use Correa et al.~\cite{Correa:2015};.
The volume integral of $\rho_{\rm host}^2$ has the closed
form~\cite{Ando:2013wia}
\begin{equation}
\int dV\,\rho_{\chi}^2(r|M) \;=\;
\frac{4\pi r_s^3\rho_s^2}{3}
\!\left[1 - \frac{1}{(1+c_{\rm vir})^3}\right].
\label{eq:rho2int}
\end{equation}

\textit{Master expression}--- 
 Combining Eqs.~(\ref{eq:HMF})
and~(\ref{eq:rho2int}) with the substructure factor
$(1+B_{\rm sh}(M,z))$ from Section~\ref{subhaloboost}~\cite{Ando:2013wia},
\begin{eqnarray}
Z_c(z) \;=\; \left(\!\frac{1}{\Omega_\chi\rho_c}\!\right)^{\!\!2}
\int_{M_{\min}}^{M_{\max}}\!\!dM\;
\frac{dn}{dM}(M,z)\;
\bigl[1 + B_{\rm sh}(M,z)\bigr]
\int dV\,\rho_{\chi}^2(r|M)
\label{eq:Zc}
\end{eqnarray}
with $M_{\min} = 10^{-6}\,M_\odot$ — the kinetic-decoupling scale
for a generic WIMP~\cite{Ullio:2002pj, Ando:2005xg} — and
$M_{\max} = 10^{15}\,M_\odot$, above the exponential HMF cutoff.
The smooth-host and substructure pieces are kept separate for
diagnostics. $Z_c \to 1$ at high $z$ where the universe has
not yet built non-linear structure, and grows toward low $z$,
with the amplitude sensitive to $M_{\min}$ through the steep
low-mass slope of the integrand. 

\textit{Implementation in $\gamma$-{\tt Cascade}.}---The interpolated
$Z_c(z)$ enters Eq.~\eqref{eq:injection} of Appendix~\ref{using gamma cascade}
directly. The comoving photon production rate fed to
\textbf{CascadeDiffuse} is
\begin{equation}
\frac{dN_\gamma}{dV\,dt\,dE_\gamma}\bigg|_{\rm src}(z)
\;=\;
\frac{1}{2}\,\langle\sigma v\rangle
\!\left(\!\frac{\rho_c\,\Omega_\chi}{m_\chi}\!\right)^{\!2}
(1+z)^6\;Z_c(z)\;
\frac{dN}{dE_\gamma}\bigg|_{E_\gamma(1+z)},
\label{eq:source}
\end{equation}
where $(1+z)^6$ accounts for the physical-density scaling of
$\bar\rho_{\rm \chi}^2(z)$ and $Z_c(z)$ from
Eq.~(\ref{eq:Zc}) provides the clumping enhancement.
$\gamma$-{\tt Cascade}~\cite{Capanema:2024nwe, Blanco:2018bbf}
propagates and absorbs the result through CMB and EBL backgrounds
as described in Appendix~\ref{using gamma cascade}. Swapping
Correa et al.~\cite{Correa:2015} for Okoli \&
Afshordi~\cite{Okoli:2016} in $c(M,z)$ shifts $Z_c(0)$ by
by a factor of around 10. The extragalactic contribution is
sub-dominant to the Galactic prompt flux for the DM masses
considered in Section~\ref{results}; we include it for
completeness.

\end{document}